\documentclass[12pt,preprint]{aastex}

\newcommand{\EGRET}{{\it EGRET}\ }

\newcommand{\et}{{\it et~al.~}}

\newenvironment{inlinefigure}{
\smallskip
\def\@captype{figure}
\noindent\begin{minipage}{0.999\linewidth}\begin{center}}
{\end{center}\end{minipage}\smallskip}

\shorttitle{}
\shortauthors{Sowards-Emmerd, Romani, Michelson \& Ulvestad}

\begin{document}

\title{Blazar Counterparts for 3EG sources at -40 $<$ decl. $<$ 0:\\
Pushing South through the Bulge}
\author{David Sowards-Emmerd\altaffilmark{1}, Roger W. Romani 
\& Peter F. Michelson\altaffilmark{1}}
\affil{Department of Physics, Stanford University, Stanford, CA 94305}
\email{dse@darkmatter.stanford.edu, rwr@astro.stanford.edu, 
peterm@stanford.edu}

\author{James S. Ulvestad}
\affil{National Radio Astronomy Observatory, Socorro, NM 87801}
\email{julvesta@nrao.edu}
\altaffiltext{1}{also, Stanford Linear Accelerator Center, 
Stanford, CA, 94039-4349}

\begin{abstract}

Supplementing existing survey data with VLA observations, we have extended 
$\gamma-$ray counterpart identifications down to decl. = -40$^\circ$ using 
our Figure of Merit approach.  We find blazar counterparts for 
$\sim$ 70\% of EGRET sources above decl. = -40$^\circ$ away from the Galaxy. 
Spectroscopic confirmation is in progress, and spectra for $\sim$ two dozen 
sources are presented here. We find evidence that increased exposure
in the bulge region allowed EGRET to detect relatively faint blazars;
a clear excess of non-blazar objects in this region however argues
for an additional (new) source class.
\end{abstract}


\section{Introduction}

The \EGRET telescope on the Compton Gamma Ray Observatory (CGRO) satellite 
has detected 271 sources in a survey of the $\gamma$-ray (~100 MeV to 10 GeV) 
sky.  Roughly one quarter of these have previously been identified as blazars 
\citep{har99,mat01}.  Recently, we have developed a new counterpart 
identification technique, which has pushed
the identified fraction to $\sim$70\% in the northern hemisphere 
(Sowards-Emmerd \et 2003, hereafter SRM03).  Here, we extend the application 
of this technique south (-40$^\circ$ $<$ decl. $<$ 0$^\circ$), including the 
Galactic bulge region.  In order to rank blazar candidates in this zone using 
our `Figure of Merit' (FoM) method, we have selected flat spectrum
radio counterpart candidates in 3EG source error boxes.
We have obtained spectroscopic confirmations and estimated redshifts for 
many of these candidate blazars with 
the Hobby-Eberly Telescope (HET) Marcario Low Resolution Spectrograph (LRS) 
 and with the  2.7-m Harlan J. Smith Telescope IGI spectrograph at
McDonald Observatory. For our final counterpart list, we used the VLA
to obtain compact
8.4 GHz fluxes similar to those previously mined from the CLASS (Meyers 
\et 2002) survey.  This allows a robust selection of EGRET blazar-like
counterparts and a statistical measure of the identification probability
of each source.

	The blazar label is somewhat heterogeneous, but in the context
of the unified AGN model, these sources are believed to be viewed close
to the axis of a powerful
relativistic jet. As such they are compact flat spectrum radio
sources, with apparent superluminal motion at VLBI scales. The
optical counterparts exhibit significant polarization and OVV (optically 
violently variable) behavior \citep{up95}. Optical spectroscopy yields a 
dichotomy of sources:  flat spectrum radio quasars with broad emission 
lines and continuum-dominated BL Lac-type objects showing weak absorption 
features and occasionally narrow, weak emission lines.  The 
broad-band spectral energy distribution (SED) is sometimes used to divide 
these into two classes, with `red' blazars showing a synchrotron peak in 
the IR-optical with a Synchrotron Self-Compton (SSC) peak in the 
$\gamma$-ray while `blue' blazars have a synchrotron component 
extending into the X-ray with a SSC peak inferred to extend to the TeV 
range (Urry 1999, and references therein).

\section{Candidate Selection and Figure of Merit}

In our previous analysis (SRM03), we quantified the correlation between 
flat spectrum radio sources and 3EG positions.  This approach produces a 
quantitative evaluation of the likelihood of each counterpart candidate, 
based on radio flux, radio spectral index, 
X-ray flux, and location of the source within a 3EG error box.
The independent functions used to generate this `Figure of Merit' (FoM)
statistic took the form of a fractional excess of sources
found within the 3EG positional 95\% confidence contours, 
in a given flux or spectral index bin, 
relative to the random background sources:
\begin{equation}
n = \frac{N_{3EG} - N_{Random}}{N_{3EG}}
\end{equation}
These functions, combined with the positional likelihood 
$L(\alpha,\delta)$, form our FoM:
\begin{equation}
FoM=n_{8.4 GHz} \times n_{\alpha} \times n_{X-ray} \times L(\alpha,\delta)
\end{equation} 
The details of the FoM analysis are described in SRM03.  Here
it is worth noting that the adopted fitting functions take the radio FoM 
to zero at $S_{8.4} =$ 85 mJy and the spectral index FoM to zero at 
$\alpha = $0.53, where $S_\nu \propto \nu^{-\alpha}$.

To generate a FoM for southern/bulge sources equivalent to those
of the Northern sample, accurate, compact 
8.4 GHz fluxes were essential.  Source selection for 8.4 GHz VLA 
A-Array observations proceeded in a similar fashion to that for the 
CLASS survey. We first selected NVSS (1.4 GHz; Condon \et 1998) 
sources within 90 arcseconds of PMN (4.85 GHz; Gregory \et 1996) 
single dish source positions, for PMN sources
in the TS maps of 3EG catalog sources in the range -40$^\circ$ $<$ decl. 
$<$ 0$^\circ$.  The `Test Statistic' TS is the maximum likelihood estimator used
in the EGRET source catalogs to quantify the probability that an EGRET 
point source is at a given location; contours of $\Delta$TS from the
maximum define the source uncertainty region.  For each 1.4~GHz source a 
spectral index was computed; if multiple NVSS sources matched one PMN source, 
the NVSS fluxes were summed and the average 1.4/4.8 GHz spectral index
was applied to each.  Sources with 
spectral indices steeper than 0.5, the CLASS spectral index cut, were 
dropped from consideration.  Additionally, a conservative 
spectral index lower limit of -2.0 was applied.  Such strongly inverted 
spectral indices are consistent with optically thick thermal emission and 
we expect these sources to be Galactic (planetary nebulae and HII regions).  
Blazars generally do not exhibit spectra nearly this inverted.  The CLASS 
blazar survey \citep{mar01} contains 6 sources (of 325) with $\alpha_{1.4/4.8} 
< -0.8$ and only one of these has been optically identified as a blazar.  
The Deep X-ray Blazar Survey \citep{per98,lan01} find only 2 sources 
(of 298) with $\alpha_{1.4/4.8} < -0.6$ and none with $\alpha_{1.4/4.8} 
< -0.8$.

Using these spectral indices, we then computed preliminary 8.4~GHz flux 
estimates.  X-ray fluxes were mined from the ROSAT All-Sky Survey bright 
and faint catalogs.  From the estimated 8.4 GHz flux, the spectral index, 
the X-ray flux, and the $\Delta$TS at the position of the radio source, 
the FoM was calculated.  Sources located at values of $\Delta$TS $>$ 13.5 
were not 
considered.  We have made a conservative cut of FoM $>$ 0.125, half 
the threshold for inclusion as 'plausible' identifications in SRM03, to allow 
for errors in the spectral index and the flux extrapolation to 8.4 GHz.  
Blazars are significantly variable and our radio survey observations are 
not simultaneous, so it is important to note this is a significant 
source of error in the spectral index estimates.  Source confusion in 
the large PMN beam and measurement uncertainties for the fainter sources 
may also contribute.  PMN observations were made in 1990, while the 
NVSS observations spanned 1993-1996.  Sources meeting this preliminary 
FoM $>$ 0.125 criterion were targeted for VLA 8.4 GHz follow-up.

\section{VLA Observations}

CLASS 8.4 GHz fluxes are in general not available below decl.$ =0^\circ$. Thus,
to establish a spectral index and flux for the sources equivalent
to that used in SRM03, sources above the preliminary FoM threshold were 
observed at 8.4 GHz with the VLA A-array during 23-25 July 2003 (Program AR517).
Observations were taken with a bandwidth of 50 MHz in each IF and typically
two 2-3 minute scans were taken per source.  In practice, a number of targets
in the original candidate list were not observed in this campaign.
A handful of EGRET error boxes at low latitude contained  
very large numbers of sources, likely due to extended Galactic emission.
These were not observed, although a number of the candidates in these
regions formally met our 1.4 GHz/4.85 GHz survey cuts. At higher latitude
a few additional sources were missed due to clerical error. In the analysis 
that follows we use the 1.4/4.85 spectral indices and extrapolated fluxes
for these objects.
Several of these sources are very likely to be thermal, but a few deserve
interferometric follow up and our classification of these should be viewed
as preliminary.  Archival 8.4 GHz fluxes were mined from the VLA Calibrator 
database 
for several of the brightest sources. Several of the target sources were
in fact used as VLA phase calibrators during the observation run. 
A total of 102 sources were observed in this campaign.

The data were edited and 
calibrated with the NRAO Astronomical Image Processing System (AIPS) 
package in the usual manner.  Maps were created, self-calibrated and 
cleaned using AIPS.  A single-component model was employed in the 
self-calibration.   Self-calibration was not performed on
a handful of the faintest (few mJy) sources.  Finally, peak intensities of 
the sources were fit.  All of the bright sources were strongly core dominated,
although about 20\% show jets on 0.1-1 arcsec scales, consistent with the blazar
classification.  The typical noise level was less than 1 mJy/beam.  Except for 
the faintest sources, the flux errors are dominated by calibration 
uncertainties at these low declinations (estimated at $\sim$3\%). 
Fluxes of sources that meet our FoM threshold appear in Table 1. Figure 1
shows the images for our highest probability new identifications. 
Precise positions and core fluxes for all observed sources are in Table 4.
The remainder of the images will be provided upon request.

Several sources showed no compact flux, and a handful gave fluxes
much lower than expected from our extrapolated values.  Indeed all but 
one of the sources with $\alpha_{1.4-4.8} <$-0.8 had little compact 
8.4 GHz flux.  Several of these sources can be identified as extended HII 
regions and planetary nebulae.  Based 
on this result, we believe that the sources classified as (G) in Table 1 
are in fact Galactic.  We note that many of these objects have spectral 
indices consistent with an optically thick thermal spectrum, 
$\alpha = -2.$  However, 
as Halpern \et (2001) have pointed out, blazars at low latitude can be 
masked in the radio by Galactic objects, as in the case of 3EG J2016+3657.  
To be conservative, we denote these as unclassified sources in our summary
Aitoff plot.

It should be noted that 1.4-8.4 GHz spectral indices are generally 
steeper than 1.4-4.8 GHz spectral indices based on the PMN fluxes.  
This is also seen in the CLASS blazar survey \citep{mar01}.

\section{Optical Follow-up}

The majority of our radio selected blazar candidates have archival 
optical classifications and redshifts.  Information for these objects 
were extracted from the SIMBAD and NED databases.  We have targeted the 
remainder of the sources for spectroscopic followup at McDonald 
Observatory. 

Spectroscopy was obtained using the Marcario LRS (Hill et al. 1998) on the 
9.2 m HET (Ramsey et al. 1998).  These targets were observed in regular queue 
operations between April 2003 and July 2003.  We obtained 2 $\times$ 300s 
exposures for most targets, and 3 $\times$ 300s for fainter targets.  
Observations were made employing a 300 line mm$^{-1}$ grating and a
2$^{\prime\prime}$ slit, 
giving a dispersion of 4\AA ~per (binned) pixel and an effective resolution 
of 16\AA ~covering $\lambda\lambda 4200-10000$\AA.  

Due to the pointing limitations of the HET (decl. $\ge$ -11$^\circ$), the 
2.7m Harlan J. Smith telescope at McDonald Observatory was used to observe 
southern targets below this limit.  Observations were made during one 
observing run from 25-29 July 2003.  The Imaging Grism Instrument (IGI) 
spectrograph was used with a 
6000\AA ~Grism, and a 50mm lens.  A 2$^{\prime\prime}$ 
slit was employed for the first 1.5 nights, and a 2.5$^{\prime\prime}$ slit
thereafter due to generally poor seeing.  The image FWHM fluctuated between 
1.5$^{\prime\prime}$ and 2$^{\prime\prime}$ on the best night, but typically 
held around 2-2.5$^{\prime\prime}$.  
The IGI setup covered a smaller wavelength range than the HET/LRS, namely 
$\lambda\lambda 4250-8500$\AA. Due 
to a wide range of conditions and source brightnesses, total exposures 
ranged from 300 to 3600 seconds.  

Standard CCD reductions were performed using IRAF.  Spectra were optimally 
extracted from the 2-d images and calibrated.  Telluric corrections were 
applied to remove atmospheric absorption.  Redshifts were estimated 
by cross-correlation analysis with AGN and galaxy spectral templates, using 
the IRAF RVSAO package.  

As of this publication, we have obtained 26 spectral classifications, 
8 with the HET/LRS 
and 18 with the McDonald 2.7m/IGI.  The new redshifts and basic optical 
properties of these objects are listed in Table 2.  Marginal (e.g. single 
line or low S/N) redshift 
estimates are denoted by a colon.  Spectra taken at McDonald Observatory
are plotted in Figure 2. Since optical observations were initiated before the
8.4 GHz data were reduced, a number of sources in Figure 2 and Table 2 do
not appear in Table 1, as their final FoM are below the `plausible' threshold.

Sources observed at McDonald Observatory were classified based on 
their spectra, S/N permitting.  The observed sources fall into the 
following 3 categories:  BL Lac objects (BLL), flat spectrum radio quasars 
(FSRQ), and narrow line radio galaxies (NLRG).  BL Lac objects are defined 
here by the following properties (Marcha et al. 1996): \\
1.  H/K break contrast $ =(\frac{f^+ - f^-}{f^+}) < 0.4$ where $f^+$ and $f^-$ 
are the fluxes redward and blueward of the break. \\
2.  Rest-frame emission line equivalent width $<$ 5\AA.\\
A handful of objects were observed to have narrow emission
lines, with kinematic widths $<$1000 km/s, and equivalent widths larger than
the BL Lac threshold.  These objects generally exhibit
a weak continuum flux, and were consequently classified as NLRGs.  
The majority of observed sources were FSRQs with broad (v$_{kin}>$ 1000-2000 
km/s) emission lines.  For the sources with archival redshifts, we examined
spectra from the literature whenever possible.  When  these were
unavailable the NED/SIMBAD classifications were adopted as a last resort.

\section{Results}

All of the 3EG error boxes located in the region -40$^\circ$ $<$ decl. $<$ 
0$^\circ$ are listed in Table 1, along with classifications and new 
counterpart candidates.  We recover many of the counterparts 
identified in the Third EGRET Catalog \citep{har99} and by Mattox \citep{mat01}.
Those that did not meet our selection criteria are included for completeness
(italics).  We also indicate in the classification column sources that are
previously discussed pulsar/plerion candidates (p). 

The spectral index listed in the table is calculated between 1.4 GHz and 
8.4 GHz, where $S_\nu \propto \nu^{-\alpha}$, with the 
exception of indices listed in parentheses which were calculated between 
1.4 GHz and 4.8 GHz.  A radio ID flag, RID, was included to 
indicate the origin of the radio flux: the VLA Calibrator Survey, 
this campaign, or both.  

Sources are divided into high and lower confidence identifications 
by their FoMs.  
Lower confidence (`plausible') identifications fall in the range 0.25 $<$FoM$<$
1.0, and higher confidence (`likely') source identifications have FoM 
$\geq$ 1.  In the -40$^\circ$ $<$ decl. $<$ 0$^\circ$ band we have identified 
30 likely and 23 plausible counterparts with 8.4 GHz flux measurements
(5 nominally `likely' counterparts lacking 8.4 GHz confirmation deserve 
follow-up, but as noted above the 28 additional sources classified as
`plausible' based on extrapolated fluxes are mostly at low $b$,
are very likely Galactic,
and are designated `G' in Table 1). Despite the rather cautious labels,
the sources with 8.4 GHz observations are statistically highly
significant identifications;
simulations based on the Northern survey indicate a success rate of 92\% 
for the likely sources and 82\% for the plausible sources. Thus among
`likely' sources with 8.4~GHz measurements (of which 7 are newly identified 
here) we statistically expect 2 false positives. For the `plausible'
sources (16 of which are new) we expect no more than 4 false identifications.

As in the north, we find evidence in this Southern extension for multiple 
counterparts for individual 3EG sources, where the fainter radio counterpart(s) 
would be 
ignored in the 3EG/Mattox studies.  Figure 3 shows two cases, 3EG J1246-0651 
and 3EG J1911-2000.  3EG J1246-0651 contains two high-confidence IDs in an 
elongated error region, with two likelihood maxima corresponding well 
to the candidate
positions.  3EG J1911-2000 also contains two counterparts, one plausible 
and one high-confidence.  Again, the sources line up along the 
major axis of the error contour.
It appears that the centroid of the uncertainty region is 
shifted slightly from the more likely source toward the fainter radio source, 
suggesting that a fainter $\gamma$-ray contribution biases
the TS localization.

\subsection{Comparison with the Northern 3EG Sample}

A comparison of the populations and luminosity functions of the candidates
requires full treatment of the EGRET sensitivities and the radio/X-ray
selection biases. We defer this population study to a future communication.
However, it is already interesting to make a comparison of the areal 
densities of various source classes in the combined survey.

The Galactic Aitoff projection in Figure 4 summarizes the current status 
of the 3EG counterpart identification.  In order to compare the different 
populations, we divide the sky above decl. $>$ -40$^\circ$ into three 
regions:  \\
-- Galactic plane -- $\vert b \vert < $10$^\circ$, \\
-- Galactic bulge (excluding the plane) -- $\vert b \vert>$10$^\circ$ and 
$<$ 30$^\circ$ from $l=b=0$, and \\ 
-- high latitude -- the remainder.  \\
In the `high latitude' set, we find 88 blazar IDs and 41 non-blazar IDs, 
suggesting
that 68\% or more of the high latitude 3EG detections have radio-bright blazar 
counterparts.  In the bulge region, we find 13 blazar IDs and 15 
non-blazar IDs, compared to 4 blazar IDs and 2 non-blazar IDs predicted 
from the areal density at high latitude.  The larger number of sources 
here, including a factor of 3 more blazars, may be attributed to deeper 
EGRET exposure (2.5$\times$ larger 3EG exposure, averaged over 
area, in our bulge region relative to the high latitude region.)  
Even relative to this increase, there is clearly an excess of 
non-blazar identifications in the bulge (54\% {\it vs.} 32\% at high latitude),
which likely represents a new 
bulge population of $\gamma-$ray sources.  Within the plane, we find 12 
blazar IDs and 49 non-blazar IDs, compared to 18 blazar IDs and 8 
non-blazars predicted from high latitude.  The apparent deficit of blazar 
identifications in the plane may be partly due to decreased EGRET sensitivity
in this high background region, although source confusion limitations
to counterpart identification may also play a role. The large number
of unidentified sources here clearly represents a Galactic population.

Grouping objects by type, we find 17\% BL Lacs and 83\% FSRQs in the Southern 
extension, very similar to the
19\% BL Lacs and 81\% FSRQs in the Northern survey.  

	We can examine the redshift distributions of the Northern and Southern
extension samples, since these should be independent of the initial
candidate selection. Before we do so, we include four additional redshifts
determined for the Northern sample since the publication of SRM03 (Figure 5,
Table 3).
Since the BL Lacs and FSRQs form two distinct populations in redshift space, 
we treat the two samples independently (Figure 6).  Similar to the Northern 
sample, we 
uncover a handful of $z>3$ sources in the Southern extension.  For the FSRQ 
sample, we find a good agreement between the north and south, with a 
Kolmogorov-Smirnov probability of 0.27 that the populations are drawn from 
the same parent distribution.  In the case of BL Lacs, the north and south 
distributions are still consistent, but with a lower probability -- 
Prob(K-S) = 0.08.  However, it should be noted that spectroscopic 
identification is much more complete in the Northern sample.

\section{Individual Object Notes}

\textit{3EG J0340-0201}--In addition to the likely association with J0339-0146
we find a plausible second counterpart in J0339-0133. 

\textit{3EG J0412-1853}--In this large error region, $\sim$4 degrees 
in diameter, 
we find a plausible counterpart, J0409-1948, in addition to the likely ID 
J0416-1851. 

\textit{3EG J0530-3626}--From the predicted radio flux and 1.4/4.8 GHz 
spectral index, we believe J0529-3555 to be the most likely counterpart,
although this requires confirmation. The 
previous claim, J0522-3627, with $\Delta$TS = 17.9 lies well to the south of 
this error map, which is strongly elongated east to west. It does not meet our
FoM cut.

\textit{3EG J0542-0655}--In addition to the previous association, J0541-0541 
(plausible), we find an equally likely counterpart based on 1.4/4.8 GHz 
spectral index and extrapolated flux; again this should be interferometrically
confirmed.

\textit{3EG J1246-0651}--We find two likely associations, J1248-0632 
and J1246-0730, in this map.  These sources present a strong case for two 
sources nearly resolved by EGRET.  The TS map is shown in Figure 3.

\textit{3EG J1500-3509}--J1457-3539 and J1505-3432, neither of which were 
previously identified as $\gamma$-ray counterparts, are both likely 
IDs located within this error map.

\textit{3EG J1504-1537}--Three plausible sources occupy this error map in 
addition to the previous ID J1507-1652, which is located well beyond the 
99\% contour.

\textit{3EG J1607-1101}--Two plausible and one likely source occupy this 
map based on extrapolated fluxes and 1.4/4.8 GHz spectral indices.  
J1605-1139 is claimed as a counterpart by Tornikoski (2002) based on 90 GHz
observations, but it does not match our survey criteria.

\textit{3EG J1612-2618}--Two likely sources are located in this error map, 
based on 1.4/4.8 GHz spectral indices and extrapolated fluxes. Two
plausible sources near the threshold for inclusion are also selected; 
these are unlikely to pass the compact 8.4~GHz FoM cut.

\textit{3EG J1718-3313}--The likely counterpart, J1717-3342, which we identify 
as a BL Lac, is very reddened at its low latitude.

\textit{3EG J1746-2851}--McLaughlin and Cordes (2003) present evidence
that PSR J1747-2958 is the counterpart.

\textit{3EG J1800-2338}--This $\gamma$-ray source has been identified with 
the pulsar PSR B1758-23. 

\textit{3EG J1809-2328}--This $\gamma$-ray source has been identified with a plerion \citep{rrk01,bra02}.

\textit{3EG J1832-2110}--counterparts J1832-2039, plausible, and J1833-2103, 
likely, share this compact, circular error box.

\textit{3EG J1911-2000}--J1911-2006, high-confidence, and J1911-1921, 
plausible, are both associations in this error box.  The $\gamma$-ray flux 
of J1911-2006 is likely supplemented by that of J1911-1921.  The TS map is 
shown in Figure 3.

\textit{3EG J1937-1529}--J1935-1602, plausible, and J1939-1525, 
likely, both lie in this elongated error box.  Our analysis does not support
the previously claimed counterpart J1941-1524.

\textit{3EG J2006-2321}--We find the centrally located J2005-2310, 
previously claimed as a new low radio flux ID by Wallace (2002), to be a 
likely counterpart.

\textit{3EG J2034-3110}--One likely ID, J2030-3039, and one plausible
ID, J2039-3157, reside in this large error box.  Neither have been previously 
associated with the EGRET source.  The 95\% contour is not closed in this map.

\section{Conclusions}

From our radio follow-up, our FoM method has selected 23 new IDs, with 8.4 
GHz fluxes, between -40$^\circ$ $<$ decl. $<$ 0$^\circ$.  Of these new
identifications, seven sources have high confidence `likely' counterparts
and the remainder are plausible.  We expect no more than 3 of these new 
sources are false positives.  A few additional identifications are based on 
4.85 GHz extrapolations and require 8.4~GHz compact fluxes for confirmation.
We also have spectroscopically identified 26 new optical counterparts 
selected for this sample.  Several of these targets fell below our FoM 
after VLA follow-up - the spectroscopic targets were selected before 
8.4~GHz fluxes were available - but are included in Table 2 for completeness.  
Combined with the Northern sample, for 
high latitude 3EG error boxes ($\vert b \vert>10^\circ$ and $>$30$^\circ$ from the 
galactic center), we find blazar IDs for 88 of 129 sources, or 68\%.
Our analysis has now been applied to 157 of the 186 3EG detections
located above $\vert b \vert>10^\circ$.

Extending the survey into the Galactic bulge, we find a surplus of 
unidentified 3EG sources. These likely
comprise a new set of Galactic $\gamma$-ray emitters.  Since the areal
density of identified blazars in the bulge region is also larger than 
in the North, it seems that 
the longer EGRET exposure in this region has let us look lower on the 
blazar luminosity function, which should be helpful in refining the 
blazar contribution to the unresolved high latitude $\gamma$-ray 
counts.  We defer detailed 
calculation of the blazar luminosity function and number count 
predictions for GLAST to a later paper.

\acknowledgments

The Hobby-Eberly Telescope (HET) is a joint project of the University 
of Texas at Austin, the Pennsylvania State University, Stanford 
University, Ludwig-Maximilians-Universit\"at M\"unchen, and 
Georg-August-Universit\"at G\"ottingen. The HET is named in honor of 
its principal benefactors, William P. Hobby and Robert E. Eberly.

The Marcario Low Resolution Spectrograph is named for Mike Marcario 
of High Lonesome Optics who fabricated several optics for the instrument 
but died before its completion. The LRS is a joint project of the 
Hobby-Eberly Telescope partnership and the Instituto de Astronomia de 
la Universidad Nacional Autonoma de Mexico.

The National Radio Astronomy Observatory is a facility of the 
National Science Foundation operated under cooperative agreement
by Associated Universities, Inc.

DSE was supported by SLAC under DOE contract
DE-AC03-76SF00515 and PFM acknowledges support from NASA contract
NAS5-00147. We thank F. Heatherington for assistance with reduction of the
VLA radio data.

\clearpage

\begin{inlinefigure}
\figurenum{1}
\scalebox{0.5}{\rotatebox{-90}{
\plotone{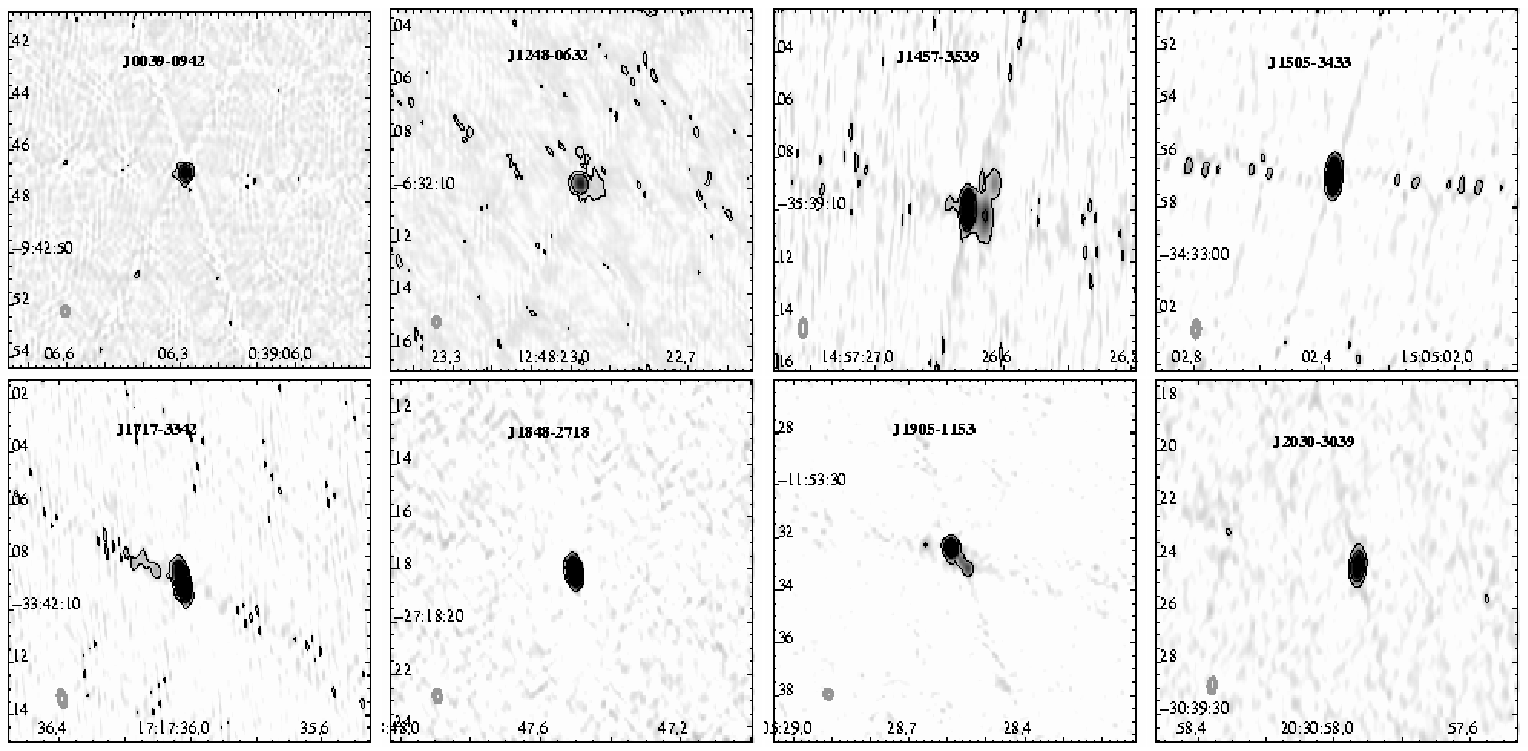}}}
\figcaption{Sample VLA 8.4~GHz snapshots of our blazar counterparts.
Here the 7 newly identified high confidence `likely' associations are shown
with contours at 1mJy and 10mJy/beam and a logarithmic grey scale to bring
out faint noise and source structure.  The restoring beam is in grey at
the lower left. One lower confidence `plausible'
association with a strong jet (J1905-1153) completes the sample set.}
\label{R1}
\end{inlinefigure}

\begin{inlinefigure}
\figurenum{2}
\scalebox{1.0}{\rotatebox{0}{
\plotone{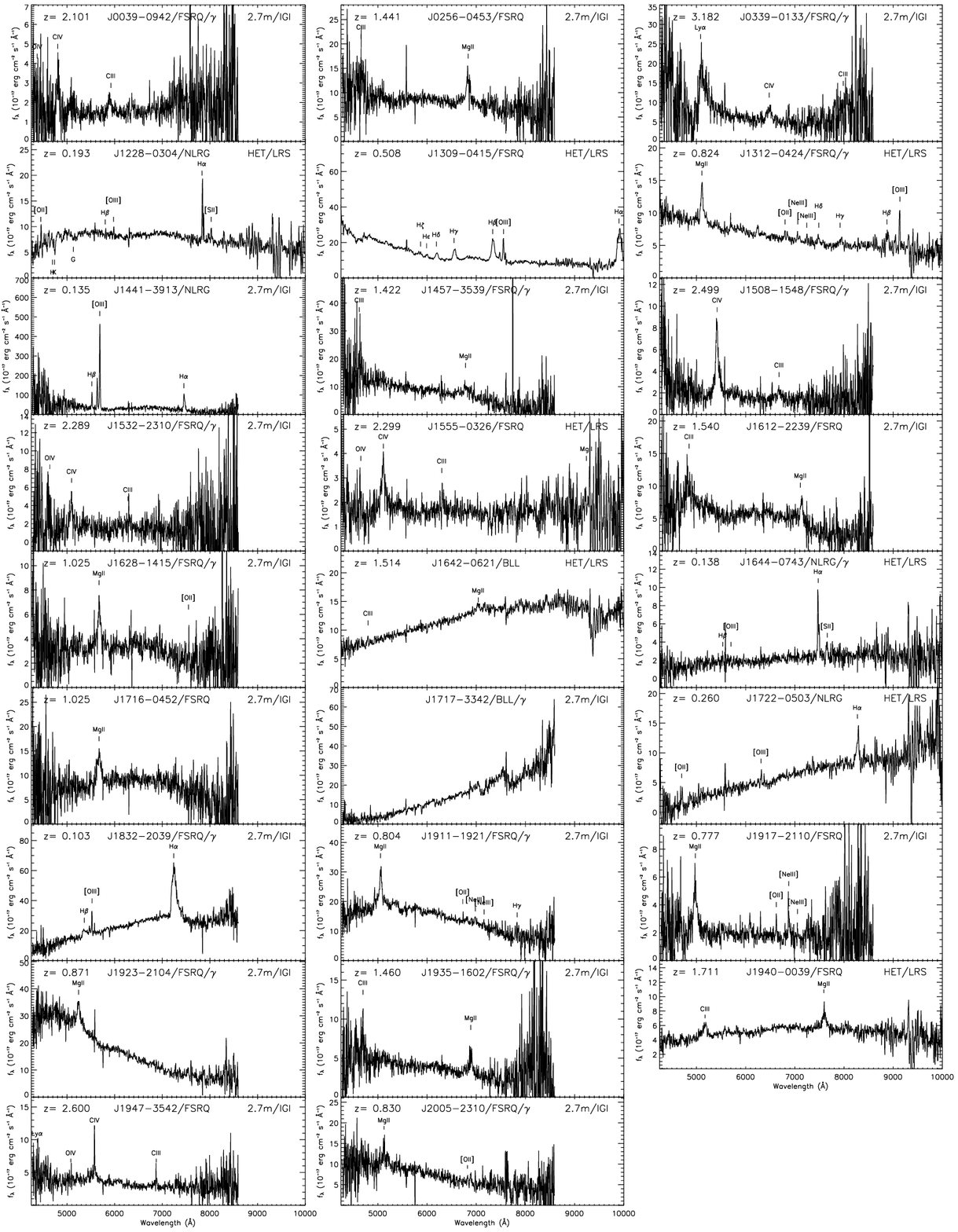}}}
\figcaption{HET/LRS and McDonald 2.7m/IGI Spectroscopy}
\label{S1}
\end{inlinefigure}

\begin{inlinefigure}
\figurenum{3}
\scalebox{.7}{\rotatebox{0}{
\plotone{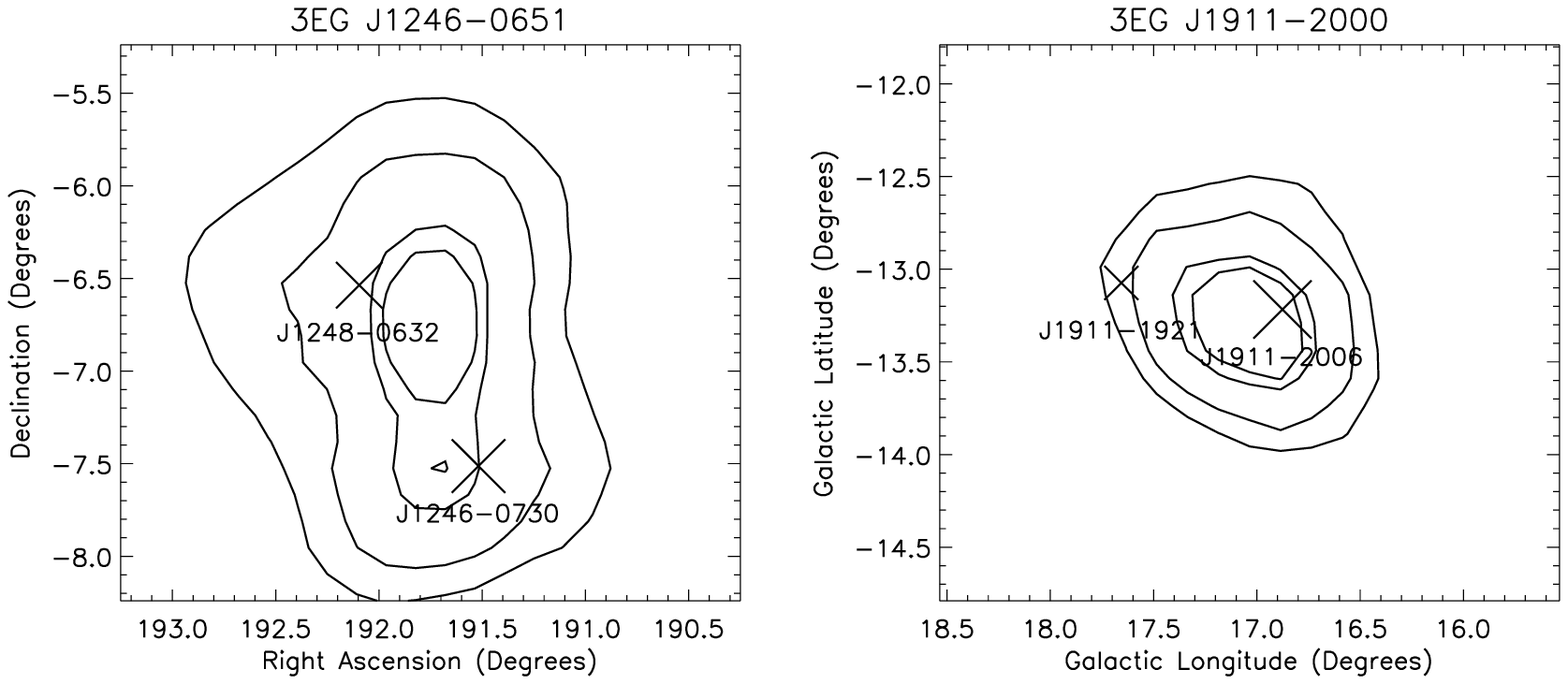}}}
\figcaption{3EG TS maps for two likely composite sources. Contours show
50\%, 68\%, 95\% and 99\% uncertainty regions. The blazar
counterparts are indicated with crosses.}
\label{TS3}
\end{inlinefigure}

\begin{inlinefigure}
\figurenum{4}
\scalebox{1.0}{\rotatebox{0}{
\plotone{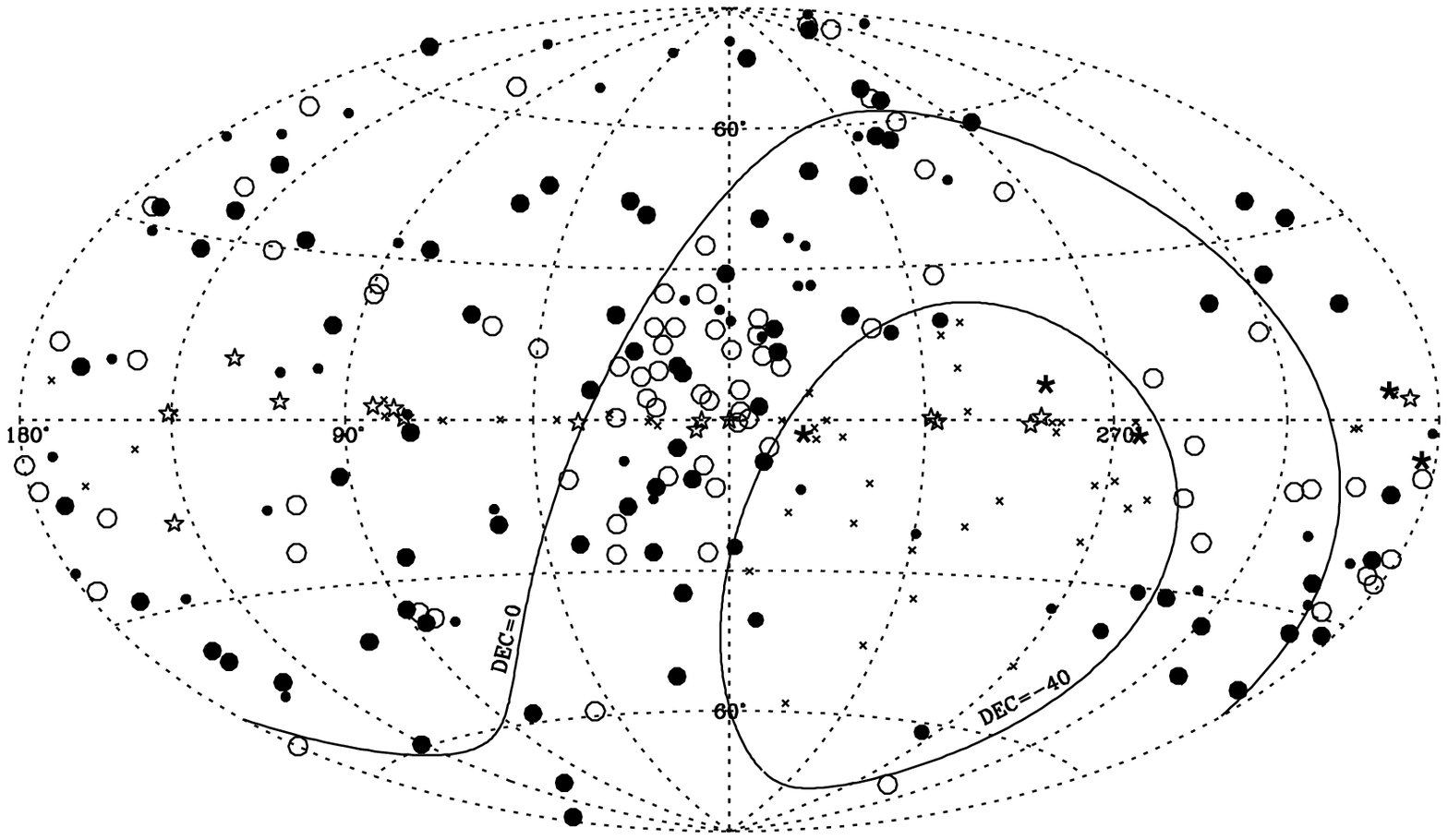}}}
\figcaption{Aitoff equal area projection of 3EG
sources in Galactic coordinates showing our new
classifications; this may be compared with Figure 4 in SRM03. 
\textit{Large filled circle}, high confidence blazar;
\textit{Smaller filled circle}, plausible blazar;
\textit{Filled star}, pulsar; 
\textit{Open star}, pulsar/plerion candidate;
\textit{Open circle}, Non-Blazar;
\textit{cross}, presently unclassified.
Symbols south of decl.=-40$^\circ$ are similar, with AGN drawn from the 3EG A/a
classifications.}
\label{A4}
\end{inlinefigure}

\begin{inlinefigure}
\figurenum{5}
\scalebox{1.0}{\rotatebox{0}{
\plotone{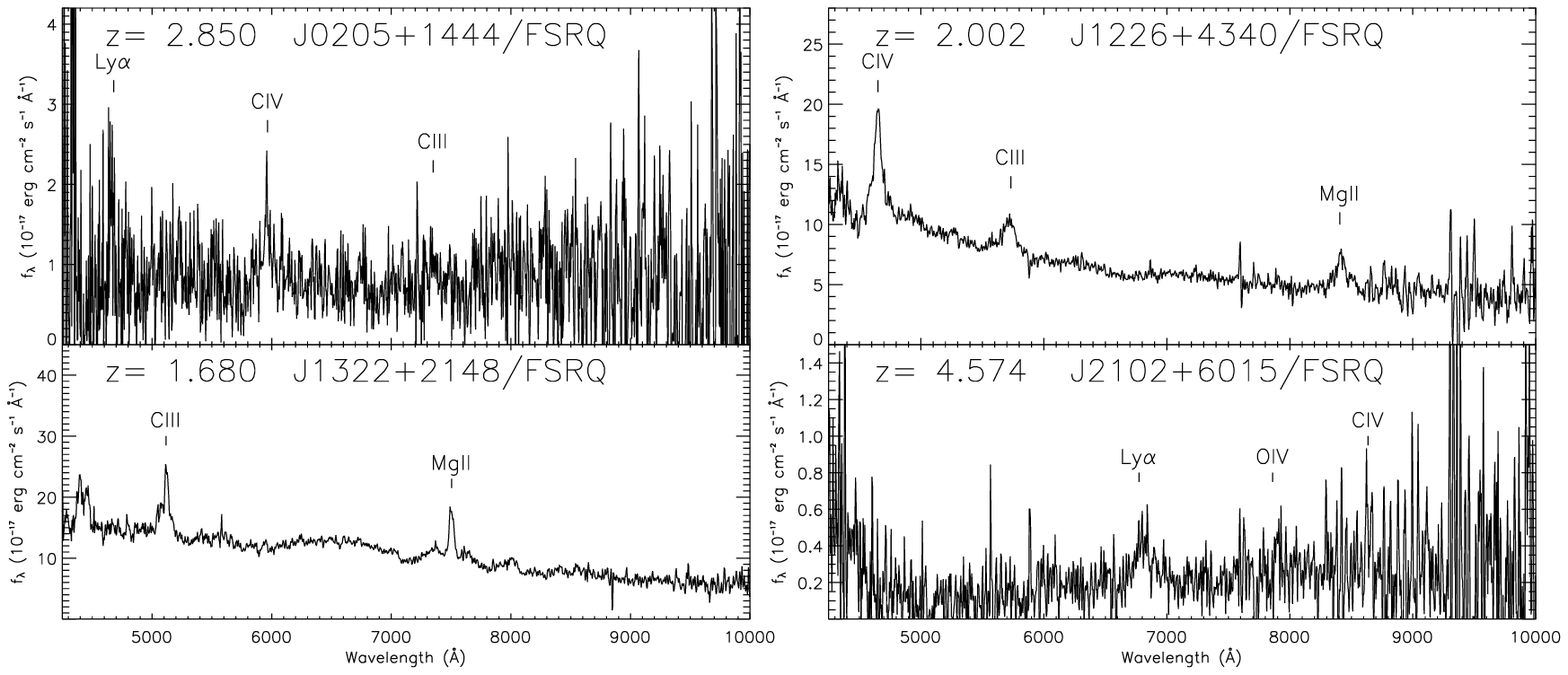}}}
\figcaption{HET/LRS spectroscopic observations of northern $\gamma$-ray source
counterparts obtained since SRM03.}
\label{S6}
\end{inlinefigure}

\begin{inlinefigure}
\figurenum{6}
\scalebox{1.0}{\rotatebox{0}{
\plotone{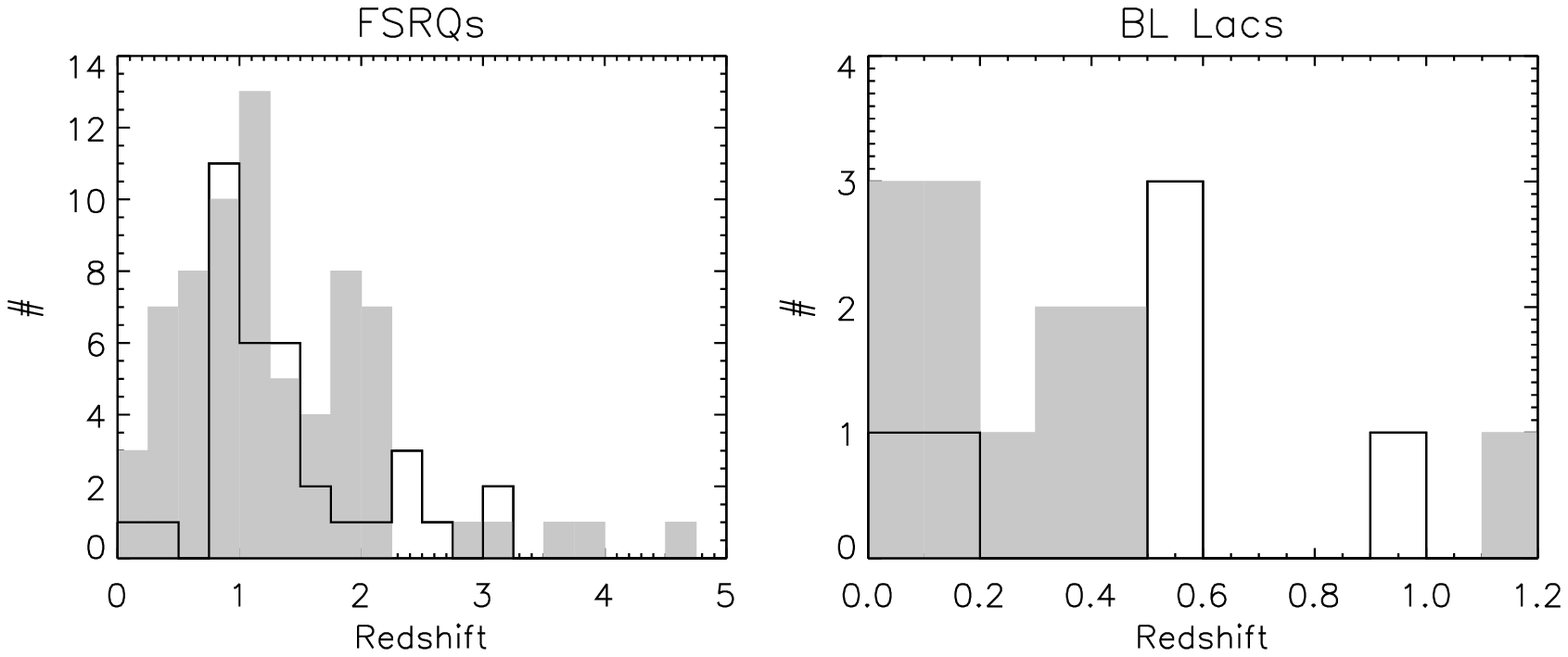}}}
\figcaption{Redshift distributions: gray histogram -- northern sources, solid line -- southern sources}
\label{zhist5}
\end{inlinefigure}

\clearpage

\begin{deluxetable}{lllllllcccccl}
\tablecolumns{13}
\tabletypesize{\tiny}
\tablewidth{0pc}
\tablecaption{3EG Objects}
\tablehead{
\colhead{3EG}&\colhead{$\gamma$-type$^a$} &\colhead{Identification} &\colhead{$S_{8.4}$(mJy)} &\colhead{$\alpha$}&\colhead{$\Delta$TS}& \colhead{FoM}& \colhead{z} & \colhead{PID$^b$}&\colhead{RID$^c$}&\colhead{Mattox$^d$}&\colhead{EGRET$^e$}&\colhead{Class$^f$}}
\startdata
J0038-0949&B&\textbf{J0039-0942}&187&-0.12&0.46&3.58&2.101&*&1&&&f\\
J0130-1758&B&\textbf{J0132-1654}&952&-0.08&7.73&1.04&1.020&&3&&a&f\\
J0159-3603&N&&&&&&&&&&&\\
J0253-0345&N&&&&&&&&&&&\\
J0340-0201&B&J0339-0133&339&0.28&6.69&0.42&3.182&$\dagger$*&1&&&f\\
&&\textbf{J0339-0146}&3010&-0.13&3.49&3.59&0.852&&3&+&A&f\\
J0412-1853
&B&J0409-1948&108&0.11&3.06&0.32&1.986&$\dagger$&1&&&f\\
&&\textbf{J0416-1851}&780&0.23&1.91&2.30&1.536&&3&--&A&f\\
J0422-0102&B&\textbf{J0423-0120}&4250&-0.2&4.81&4.03&0.915&&2&+&A&f\\
J0442-0033&b&J0442-0017&860&0.39&4.56&0.54&0.844&&2&+&A&f\\
J0456-2338&B&\textbf{J0457-2324}&1850&-0.0&0.95&7.42&1.003&&2&+&A&f\\
J0500-0159&B&\textbf{J0501-0159}&3700&-0.25&0.86&10.60&2.286&&2&+&A&f\\
J0530-3626&B&\textbf{J0529-3555}&(289)&(-0.42)&2.29&(4.07)&--&(*)&0&&&\\
&&\textit{J0522-3627}&5000&0.63&17.9&0.00&0.061&&2&&a&b\\
J0531-2940
&b&J0539-2839&760&0.06&9.83&0.32&3.104&&2&&a&f\\
J0542-0655&b&J0541-0541&980&-0.06&10.33&0.34&0.839&&2&&a&f\\
&&J0545-0539&(151)&(0.11)&6.48&(0.34)&--&&0&&&\\
J0616-0720&N&&&&&&&&&&&\\
J0616-3310&N&&&&&&&&&&&\\
J0622-1139&N&\textit{J0619-1140}&210&0.47&3.76&0.13&--&&1&--&A&\\
J0706-3837&N&&&&&&&&&&&\\
J0747-3412&N&&&&&&&&&&&\\
J0812-0646&N&\textit{J0808-0751}&2600&-0.29&14.19&0.05&1.837&&2&&a&f\\
J0852-1216
&B&\textbf{J0850-1213}&625&-0.37&2.67&5.04&0.57&&1&--&A&b\\
J0903-3531&N&&&&&&&&&&&\\
J1134-1530&N&\textit{J1130-1449}&3060&0.30&18.71&0.01&1.187&&2&&a&f\\
J1219-1520&b&J1222-1645&145&0.03&7.73&0.25&--&*&1&&&\\
J1230-0247
&N&\textit{J1232-0224}&629&0.50&3.25&0.15&1.045&&1&--&A&f\\
J1234-1318&N&&&&&&&&&&&\\
J1246-0651&B&\textbf{J1248-0632}&353&0.02&3.21&1.83&0.762&$\dagger$&1&&&f\\
&&\textbf{J1246-0730}&630&-0.08&2.09&4.05&1.286&&2&--&A&f\\ 
J1255-0549&B&\textbf{J1256-0547}&15600&-0.2&6.99&3.65&0.538&&2&+&A&b\\
J1310-0517
&b&J1312-0424&286&-0.15&7.85&0.67&0.824&*&1&&&f\\
J1314-3431
&b&J1316-3338&1100&0.07&12.5&0.14&1.210&&1&&a&f\\
J1339-1419
&B&\textbf{J1337-1257}&3850&-0.21&5.75&3.38&0.539&&3&+&A&b\\
J1409-0745&B&\textbf{J1408-0752}&630&0.00&3.50&2.19&1.494&&2&+&A&f\\
J1447-3936&N&&&&&&&&&&&\\
J1457-1903&b&J1459-1810&158&-0.24&9.82&0.25&--&*&1&&&\\
J1500-3509&B&\textbf{J1457-3539}&606&0.04&4.27&1.76&1.422&$\dagger$*&1&&&f\\
&&\textbf{J1505-3432}&393&-0.59&4.32&3.37&--&*&1&&&\\
J1504-1537&b&J1502-1508&133&-0.18&5.25&0.62&--&*&1&&&\\
&&J1505-1610&130&-0.24&5.43&0.61&--&*&1&&&\\
&&J1508-1548&307&-0.03&8.50&0.41&2.499&*&1&&&f\\
&&\textit{J1507-1652}&2400.&0.05&13.97&0.03&0.876&&2&&a&f\\
J1512-0849&B&\textbf{J1512-0905}&2150&0.12&1.79&4.31&0.360&&2&+&A&f\\
J1517-2538
&b&J1517-2422&1930&0.02&9.18&0.44&0.049&&3&--&a&b\\
J1527-2358
&b&J1532-2310&158&-0.09&5.82&0.64&2.289&*&1&&&f\\
J1600-0351&N&&&&&&&&&&&\\
J1607-1101&B&J1603-1007&(146)&(0.08)&6.75&(0.32)&--&&0&&&\\
&&J1605-1012&(127)&(-0.27)&4.15&(0.76)&--&&0&&&\\
&&\textbf{J1612-1133}&(169)&(-0.58)&4.91&(1.52)&--&(*)&0&&&\\
&&\textit{J1605-1139$^g$}&&&&&&&&&&\\
J1612-2618&B&J1607-2656&(119)&(0.29)&2.55&(0.30)&--&&0&&&\\
&&\textbf{J1611-2612}&(113)&(-0.17)&0.32&(1.54)&--&(*)&0&&&\\
&&J1613-2750&(210)&(0.05)&8.54&(0.26)&--&&0&&&\\
&&\textbf{J1617-2537}&(153)&(-0.15)&3.06&(1.22)&--&(*)&0&&&\\
J1616-2221&N&&&&&&&&&&&\\
J1625-2955&B&\textbf{J1626-2951}&2250&-0.00&4.06&2.58&0.815&&2&+&A&f\\
J1626-2519&b&J1625-2527&1100&0.45&0.65&0.99&0.786&&2&+&A&f\\
J1627-2419&N&&&&&&&&&&&\\
J1631-1018&N&&&&&&&&&&&\\
J1633-3216&N&&&&&&&&&&&\\
J1634-1434&b&J1628-1415&275&0.14&6.85&0.54&1.025&$\dagger$*&1&&&f\\
J1635-1751&b&J1629-1720&(245)&(-0.56)&9.70&(0.39)&--&&0&&&\\
J1638-2749&N&&&&&&&&&&&\\
J1646-0704
&b&J1644-0743&129&-0.08&4.73&0.55&0.139&*&1&&&r\\
J1649-1611&N&&&&&&&&&&&\\
J1652-0223&N&&&&&&&&&&&\\
J1653-2133&N&&&&&&&&&&&\\
J1709-0828&N&&&&&&&&&&&\\
J1714-3857&G&J1711-3908&(25100.)&(-1.98)&9.49&(0.71)&--&&0&&&(G)\\
&&J1712-3840&(468)&(-0.98)&4.87&(3.34)&--&&0&&&(G)\\
&&J1713-3900&(4340)&(-1.06)&0.00&(19.36)&--&&0&&&(G)\\
&&J1714-3810&(10100)&(-0.84)&9.06&(0.94)&--&&0&&&(G)\\
&&J1714-3937&(165)&(-1.31)&9.43&(0.28)&--&&0&&&(G)\\
&&J1714-3925&(203)&(-1.76)&6.56&(1.31)&--&&0&&&(G)\\
&&J1715-3913&(4300)&(-1.34)&1.81&(11.08)&--&&0&&&(G)\\
J1717-2737&N&&&&&&&&&&&\\
J1718-3313&B&\textbf{J1717-3342}&685&-0.06&2.87&3.18&--&*&1&&&b\\
J1719-0430&N&&&&&&&&&&&\\
J1726-0807&N&&&&&&&&&&&\\
J1733-1313&B&\textbf{J1733-1304}&5110&0.07&3.90&2.86&0.902&&3&+&A&b\\
J1734-3232&N&&&&&&&&&&&\\
J1735-1500&B&\textbf{J1738-1503}&(247)&(0.15)&3.28&(1.06)&--&(*)&0&&&\\
J1736-2908&N&&&&&&&&&&&\\
J1741-2050&N&&&&&&&&&&&\\
J1741-2312&N&&&&&&&&&&&\\
J1744-0310&B&\textbf{J1743-0350}&6240&-0.84&3.99&5.98&1.054&&3&+&A&f\\
J1744-3011&N&&&&&&&&&&&\\
J1744-3934&N&&&&&&&&&&&\\
J1746-1001&N&&&&&&&&&&&\\
J1746-2851&N&PSR J1747-2958 &&&&&&&&&&p\\
J1757-0711&N&&&&&&&&&&&\\
J1800-0146&N&&&&&&&&&&&\\
J1800-2338&N&PSR B1758-23&&&&&&&&&&p\\
J1800-3955
&B&\textbf{J1802-3940}&1700&0.14&2.59&2.74&--&&1&--&A&\\
J1809-2328&N&RRK&&&&&&&&&&p\\
J1810-1032&N&&&&&&&&&&&\\
J1812-1316&N&&&&&&&&&&&\\
J1823-1314&G&J1822-1251&(753)&(-1.39)&12.48&(0.32)&--&&0&&&(G)\\
&&J1824-1251&(928)&(-1.01)&11.61&(0.46)&--&&0&&&(G)\\
&&J1825-1315&(15300)&(-1.33)&8.48&(1.36)&--&&0&&&(G)\\
J1824-1514&G&J1819-1501&(2260)&(-0.80)&13.25&(0.27)&--&&0&&&(G)\\
&&J1819-1530&(982)&(-1.28)&10.31&(0.66)&--&&0&&&(G)\\
&&J1821-1449&(6150)&(-1.98)&10.87&(0.66)&--&&0&&&(G)\\
&&J1822-1550&(316)&(-1.88)&10.00&(0.45)&--&&0&&&(G)\\
&&J1822-1602&(342)&(-1.98)&10.43&(0.43)&--&&0&&&(G)\\
&&J1825-1449&(2550)&(-1.05)&4.00&(5.62)&--&&0&&&(G)\\
&&J1826-1600&(269)&(-0.99)&10.00&(0.41)&--&&0&&&(G)\\
J1826-1302&G&J1824-1251&(918)&(-1.01)&6.32&(3.00)&--&&0&&&(G)\\
&&J1825-1315&(15100)&(-1.33)&3.82&(5.24)&--&&0&&&(G)\\
&&J1826-1338&(245)&(-1.26)&6.60&(1.53)&--&&0&&&(G)\\
J1832-2110&B&J1832-2039&1070&0.03&8.18&0.70&0.103&*&1&&&f\\
&&\textbf{J1833-2103}&6750.&0.25&4.31&1.29&2.510&&2&+&A&f\\
J1834-2803&N&&&&&&&&&&&\\
J1837-0423&G&J1837-0508&(276)&(-1.92)&10.74&(0.35)&--&&0&&&(G)\\
&&J1839-0419&(1890)&(-1.60)&5.43&(4.19)&--&&0&&&PN\\
J1837-0606&N&&&&&&&&&&&\\
J1847-3219&N&&&&&&&&&&&\\
J1850-2652&B&\textbf{J1848-2718}&417&-1.26&2.38&5.85&--&*&1&&&\\
J1858-2137&N&&&&&&&&&&&\\
J1904-1124&b&J1905-1153&154&0.26&3.09&0.48&--&*&1&&&\\
J1911-2000&B&\textbf{J1911-2006}&2190&0.11&0.12&6.75&1.119&&3&+&A&f\\
&&J1911-1921&182&0.06&6.84&0.45&0.804&*&1&&&f\\
J1921-2015&b
&J1923-2104&2190&0.19&8.94&0.32&0.871&&1&--&a&f\\
J1937-1529&B&J1935-1602&286&0.04&6.68&0.73&1.460&*&1&&&f\\
&&\textbf{J1939-1525}&670&-0.0&1.69&4.75&1.657&&2&--&A&f\\
&&\textit{J1941-1524}&&&&&0.452&&&--&&r\\
J1940-0121&N&&&&&&&&&&&\\
J1949-3456&N&&&&&&&&&&&\\
J1955-1414&N&&&&&&&&&&&\\
J2006-2321&B&\textbf{J2005-2310$^h$}&230&0.14&0.36&2.70&0.830&&1&--&&f\\
J2020-1545&N&&&&&&&&&&&\\
J2025-0744&B&\textbf{J2025-0735}&727&0.33&0.89&2.20&1.388&&1&--&A&f\\
J2034-3110&B&\textbf{J2030-3039}&229&-0.05&5.45&1.00&--&*&1&&&\\
&&J2039-3157&231&0.06&6.24&0.67&--&*&1&&&\\
J2158-3023&B&\textbf{J2158-3013}&204&0.44&0.36&1.21&0.117&&1&--&A&b\\
J2251-1341&N&&&&&&&&&&&\\
J2321-0328
&B&\textbf{J2323-0317}&918&-0.00&1.97&5.12&1.411&&3&--&A&f\\
\enddata

\tablecomments{Our high confidence associations are listed in boldface and our lower confidence associations in plain text.  Previously claimed AGN associations not supported by our analysis are given in italics.  () entries are PREDICTED values extrapolating from 1.4/4.8 GHz to 8.4 GHz.}

\tablenotetext{a}{$\gamma-$ray source classification: N = non-blazar, b = plausible blazar ID, B = high confidence blazar ID, G = likely galactic}
\tablenotetext{b}{New associations and/or redshifts: asterisk indicates new spectral identification, dagger indicates archival classification.}
\tablenotetext{c}{Radio data origin: 0 = not observed, 1 = this VLA campaign, 2 = VLA Calibrator survey, 3 = both, fluxes from this VLA campaign are listed.}
\tablenotetext{d}{Mattox \et 2001 selected blazars: + = high probability, - = plausible}
\tablenotetext{e}{Third EGRET Catalog blazars: A = high confidence, a = lower confidence}
\tablenotetext{f}{Classification: b = BL Lac , f = FSRQ, r = NLRG, p = pulsar/plerion, G = likely galactic, PN = Planetary Nebula}
\tablenotetext{g}{Tornikoski \et 2002}
\tablenotetext{h}{Wallace \et 2002}

%$^a$ $\gamma-$ray source classification: N = non-blazar, b = plausible blazar ID, B = high confidence blazar ID, G = likely galactic\\
%$^b$ New associations and/or redshifts: asterisk indicates new spectral identification, dagger indicates archival classification.\\
%$^c$ Radio data origin: 0 = not observed, 1 = this VLA campaign, 2 = VLA Calibrator survey, 3 = both, fluxes from this VLA campaign are listed.\\
%$^d$ Mattox \et 2001 selected blazars: + = high probability, - = plausible\\
%$^e$ Third EGRET Catalog blazars: A = high confidence, a = lower confidence\\
%$^f$ Classification: b = BL Lac , f = FSRQ, r = NLRG, p = pulsar/plerion, G = likely galactic, PN = Planetary Nebula\\
%$^g$ Tornikoski \et 2002\\
%$^h$ Wallace \et 2002\\
%() entries are PREDICTED values extrapolating from 1.4/4.8 GHz to 8.4 GHz

\end{deluxetable}

\clearpage

\begin{deluxetable}{lllllllll}
\tabletypesize{\scriptsize}
\tablecaption{HET/LRS and McDonald 2.7m/IGI Spectroscopy}
\tablewidth{0pc}
\tablehead{
\colhead{Name} &\colhead{FoM} &\colhead{R.A. (J2000)} & \colhead{Dec. (J2000)} & \colhead{R2$^a$} & \colhead{B2$^a$} & \colhead{z} & \colhead{Type$^b$}
}
\startdata
J0039-0942&3.58&00 39 06.28&-09 42 47.5&...&20.6&2.1015&f\\
J0256-0453&0.02&02 56 50.35&-04 53 43.7&18.3&19.9&1.4410&f\\
J0339-0133&0.42&03 39 00.94&-01 33 18.1&&&3.1826&f\\
J1228-0304&0.20&12 28 36.91&-03 04 39.3&19.7&20.2&0.1930&r\\
J1309-0415&0.00&13 09 17.81&-04 15 15.2&18.8&19.0&0.5089&f\\
J1312-0424&0.67&13 12 50.92&-04 24 50.7&19.3&18.7&0.8249&f\\
J1441-3913&0.00&14 41 15.13&-39 13 48.4&14.9&15.2&0.1356&r\\
J1457-3539&1.76&14 57 26.76&-35 39 09.2&18.9&17.8&1.4222:&f\\
J1508-1548&0.41&15 08 35.69&-15 48 31.6&19.2&20.2&2.4990&f\\
J1532-2310&0.64&15 32 31.53&-23 10 32.2&19.7&21.4&2.2896&f\\
J1555-0326&0.16&15 55 30.80&-03 26 49.4&19.7&20.7&2.2996&f\\
J1612-2239&0.02&16 12 28.48&-22 39 46.8&19.5&19.6&1.5404&f\\
J1628-1415&0.54&16 28 46.71&-14 15 41.9&...&20.4&1.0255&f\\
J1642-0621&0.00&16 42 02.22&-06 21 23.6&19.2&20.9&1.5143:&b\\
J1644-0743&0.55&16 44 52.03&-07 43 43.4&...&...&0.1389&r\\
J1717-3342&3.18&17 17 36.01&-33 42 08.1&...&...&...&b\\
J1716-0452&0.12&17 16 26.57&-04 52 12.5&18.7&20.6&1.0258&f\\
J1722-0503&0.08&17 22 03.61&-05 03 25.8&18.5&...&0.2606&r\\
J1832-2039&0.70&18 32 11.10&-20 39 47.9&...&20.2&0.1033&f\\
J1911-1921&0.45&19 11 56.50&-19 21 52.0&18.3&18.5&0.8046&f\\
J1917-2110&0.13&19 17 08.66&-21 10 31.2&...&21.5&0.7775&f\\
J1923-2104&0.32&19 23 32.20&-21 04 33.8&...&...&0.8719&f\\
J1935-1602&0.73&19 35 35.79&-16 02 32.3&19.3&20.1&1.4602&f\\
J1940-0039&0.01&19 40 09.00&-00 39 01.3&17.0&18.1&1.7110&f\\
J1947-3542&0.00&19 47 22.65&-35 42 04.2&19.2&19.9&2.600:&f\\
J2005-2310&2.70&20 05 56.61&-23 10 28.1&18.8&19.6&0.8301:&f\\
\enddata
\tablenotetext{a}{R2 and B2 are USNO B1.0 optical magnitudes \citep{mon03}.}
\tablenotetext{b}{Classification: b = BL Lac , f = FSRQ, r = NLRG}
\end{deluxetable}

\clearpage

\begin{deluxetable}{llllllll}
\tabletypesize{\scriptsize}
\tablecaption{HET Spectroscopy: Northern Followup}
\tablewidth{0pc}
\tablehead{
\colhead{Name} &\colhead{FoM} &\colhead{R.A. (J2000)} & \colhead{Dec. (J2000)} & \colhead{R2$^a$} & \colhead{B2$^a$} & \colhead{z} & \colhead{Type$^b$}
}
\startdata
J0205+1444&2.58&02 05 13.12&+14 44 32.4&...&...&2.8504&f\\
J1226+4340&0.95&12 26 57.91&+43 40 58.4&19.2&19.3&2.0023&f\\
J1322+2148&0.29&13 22 11.40&+21 48 12.3&19.4&19.3&1.6803&f\\
J2102+6015&0.41&21 02 40.22&+60 15 09.8&...&...&4.5749:&f\\
\enddata
\tablenotetext{a}{R2 and B2 are USNO B1.0 optical magnitudes \citep{mon03}.}
\tablenotetext{b}{Classification: f = FSRQ}
\end{deluxetable}

\begin{deluxetable}{lllll}
\tablecolumns{5}
\tablewidth{0pc}
\tabletypesize{\scriptsize}
\tablecaption{VLA 'A' Array 8.4 GHz Observations}
\tablehead{
\colhead{Name} & \colhead{Flux [mJy]} & \colhead{R. A. (J2000)} & 
\colhead{Decl. (J2000)} & \colhead{Class}}
\startdata
J0039-0942&187  $\pm$ 6&00 39 06.291&-09 42 46.88&\\
J0128-1722&117  $\pm$ 4&01 28 01.393&-17 22 24.25&Jet?\\
J0132-1654&952  $\pm$ 29&01 32 43.487&-16 54 48.52&\\
J0239-0234&595  $\pm$ 18&02 39 45.472&-02 34 40.91&\\
J0256-0453&87.4 $\pm$ 2.6&02 56 50.420&-04 53 43.50&\\
J0339-0133&339  $\pm$ 10&03 39 00.985&-01 33 17.60&\\
J0339-0146&3010 $\pm$ 90&03 39 30.937&-01 46 35.80&\\
J0405-1906&31.0 $\pm$ 0.9&04 05 49.678&-19 06 57.27&\\
J0409-1948&108  $\pm$ 3&04 09 40.549&-19 48 01.78&\\
J0416-1851&779  $\pm$ 23&04 16 36.544&-18 51 08.34&Jet?\\
J0522-2955&47.3 $\pm$ 1.6&05 23 00.131&-29 55 17.36&\\
J0527-3708&0.&&&\\
J0531-3533&122  $\pm$ 4&05 31 30.419&-35 33 32.65&\\
J0615-3355&121  $\pm$ 4&06 15 12.689&-33 55 53.07&\\
J0619-1140&210  $\pm$ 6&06 19 04.103&-11 40 54.89&\\
J0621-1259&0.&&&PN\\
J0625-1133&183  $\pm$ 6&06 25 49.339&-11 33 24.37&Jet?\\
J0703-3746&0.&&&\\
J0710-3813&18.9 $\pm$ 1.4&07 10 47.043&-38 13 45.72&\\
J0713-3812&23.8 $\pm$ 0.9&07 13 01.858&-38 12 28.22&\\
J0747-3310&131  $\pm$ 9&07 47 19.693&-33 10 47.03&\\
J0754-3448&0.&&&\\
J0848-1159&42.5 $\pm$ 1.3&08 48 47.491&-11 59 53.68&\\
J0850-1213&625  $\pm$ 19&08 50 09.634&-12 13 35.39&Jet?\\
J0856-1105&357  $\pm$ 11&08 56 41.805&-11 05 14.42&Jet?\\
J1222-1645&145  $\pm$ 4&12 22 16.097&-16 45 54.89&\\
J1223-1544&61.3 $\pm$ 1.8&12 23 12.272&-15 44 06.63&Jet?\\
J1228-0304&128  $\pm$ 4&12 28 36.916&-03 04 39.32&\\
J1231-1236&80.6 $\pm$ 2.4&12 31 50.265&-12 36 37.04&\\
J1232-0224&629  $\pm$ 19&12 32 00.014&-02 24 04.80&Jet\\
J1246-0730&682  $\pm$ 20&12 46 04.232&-07 30 46.57&\\
J1248-0632&353  $\pm$ 11&12 48 22.975&-06 32 09.80&Jet?\\
J1308-0500&0.&&&\\
J1309-0415&84.7 $\pm$ 2.5&13 09 17.831&-04 15 16.20&\\
J1312-0424&286  $\pm$ 9&13 12 50.900&-04 24 49.88&\\
J1316-3338&1100 $\pm$ 30&13 16 07.985&-33 38 59.17&Jet?\\
J1316-3429&0.&&&\\
J1332-1402&154  $\pm$ 5&13 32 30.929&-14 02 13.16&Jet?\\
J1332-1256&99.4 $\pm$ 3.0&13 32 39.251&-12 56 15.34&\\
J1337-1257&3850 $\pm$ 120&13 37 39.782&-12 57 24.69&\\
J1441-3913&68.9 $\pm$ 2.1&14 41 15.088&-39 13 48.03&\\
J1443-3908&36.7 $\pm$ 1.1&14 43 57.197&-39 08 39.73&\\
J1457-3539&606  $\pm$ 18&14 57 26.711&-35 39 09.98&Jet\\
J1459-1810&158  $\pm$ 5&14 59 28.763&-18 10 45.19&\\
J1502-1508&133  $\pm$ 4&15 02 25.017&-15 08 52.50&\\
J1505-1610&130  $\pm$ 4&15 05 22.950&-16 10 40.60&\\
J1505-3432&393  $\pm$ 12&15 05 02.371&-34 32 56.83&\\
J1507-1652&2090 $\pm$ 60&15 07 04.786&-16 52 30.26&Jet?\\
J1508-1548&307  $\pm$ 9&15 08 35.700&-15 48 31.51&\\
J1513-2558&72.7 $\pm$ 2.2&15 13 53.174&-25 58 30.16&Jet\\
J1517-2618&92.2 $\pm$ 2.8&15 17 26.621&-26 18 18.99&\\
J1517-2422&1930 $\pm$ 60&15 17 41.813&-24 22 19.47&Jet\\
J1530-2410&55.6 $\pm$ 1.7&15 30 17.018&-24 10 46.41&\\
J1532-2310&158  $\pm$ 5&15 32 31.529&-23 10 32.43&Jet?\\
J1555-0326&232  $\pm$ 7&15 55 30.748&-03 26 49.51&Jet?\\
J1612-2239&107  $\pm$ 3&16 12 28.444&-22 39 46.68&\\
J1626-2951&1140 $\pm$ 30&16 26 06.020&-29 51 26.97&\\
J1627-2426&71.2 $\pm$ 2.1&16 27 00.008&-24 26 40.44&\\
J1627-0939&2.95 $\pm$ 0.15&16 27 45.453&-09 39 45.58&\\
J1628-1415&275  $\pm$ 8&16 28 46.618&-14 15 41.82&\\
J1632-1052&140  $\pm$ 4&16 32 50.108&-10 52 31.94&\\
J1634-1440&0.&&&\\
J1639-0715&14.4 $\pm$ 0.4&16 39 21.996&-07 15 33.12&\\
J1644-0750&1.12 $\pm$ 0.12&16 44 00.768&-07 50 00.54&\\
J1644-0743&129  $\pm$ 4&16 44 52.058&-07 43 43.10&\\
J1653-0150&68.3 $\pm$ 2.1&16 53 57.800&-01 49 59.32&\\
J1716-0452&436  $\pm$ 13&17 16 26.487&-04 52 11.94&\\
J1717-3342&685  $\pm$ 21&17 17 36.028&-33 42 08.91&Jet\\
J1719-2720&27.0 $\pm$ 0.8&17 19 59.708&-27 20 42.60&\\
J1721-2711&0.&&&PN\\
J1722-0503&310  $\pm$ 9&17 22 03.539&-05 03 25.00&Jet\\
J1729-0735&103  $\pm$ 3&17 29 34.946&-07 35 32.38&\\
J1733-1304&5110 $\pm$ 150&17 33 02.705&-13 04 49.54&Jet?\\
J1734-3314&0.&&&\\
J1740-1515&55.5 $\pm$ 1.7&17 40 03.344&-15 15 52.89&\\
J1743-0350&6240 $\pm$ 190&17 43 58.856&-03 50 04.61&\\
J1800-3849&46.1 $\pm$ 1.4&18 00 11.821&-38 49 52.50&Extended\\
J1802-3940&1700 $\pm$ 50&18 02 42.677&-39 40 07.90&Jet?\\
J1802-0207&1.36 $\pm$ 0.14&18 02 49.678&-02 07 49.15&\\
J1832-2039&1070 $\pm$ 30&18 32 11.047&-20 39 48.20&Jet\\
J1837-0629&0.&&&SNR025.5+00.2\\
J1837-0616&0.&&&\\
J1848-2718&417  $\pm$ 13&18 48 47.504&-27 18 18.09&\\
J1850-2740&90.0 $\pm$ 2.7&18 50 13.979&-27 40 20.92&\\
J1905-1153&154  $\pm$ 5&19 05 28.591&-11 53 32.38&Jet\\
J1911-2006&2190 $\pm$ 70&19 11 09.652&-20 06 55.10&Jet?\\
J1911-1921&182  $\pm$ 5&19 11 56.517&-19 21 50.97&Jet\\
J1917-2110&195  $\pm$ 6&19 17 08.642&-21 10 30.78&\\
J1923-2104&2200 $\pm$ 70&19 23 32.189&-21 04 33.33&\\
J1935-1602&286  $\pm$ 9&19 35 35.795&-16 02 32.38&\\
J1940-0039&89.1 $\pm$ 2.7&19 40 09.000&-00 39 02.01&Jet\\
J1947-3542&33.5 $\pm$ 1.0&19 47 22.624&-35 42 03.69&\\
J1952-3412&116  $\pm$ 3&19 52 00.109&-34 12 26.45&Jet\\
J2005-2310&230  $\pm$ 7&20 05 56.594&-23 10 27.01&\\
J2025-0735&727  $\pm$ 22&20 25 40.659&-07 35 52.70&Jet\\
J2030-3039&229  $\pm$ 7&20 30 57.931&-30 39 24.36&\\
J2039-3157&231  $\pm$ 7&20 39 08.653&-31 57 03.90&\\
J2158-3013&204  $\pm$ 6&21 58 52.064&-30 13 32.10&Jet\\
J2318-0352&81.2 $\pm$ 2.4&23 18 15.623&-03 52 14.85&Jet?\\
J2323-0150&233  $\pm$ 7&23 23 04.629&-01 50 48.10&\\
J2323-0317&918  $\pm$ 28&23 23 31.953&-03 17 05.02&Jet?\\
J2326-0202&226  $\pm$ 7&23 26 53.776&-02 02 13.76&Jet\\
\enddata
\tablecomments{Peak 8.4 GHz flux densities and positions are based on 2-d gaussian (IMFIT) fits to the bright core.  Estimated 3\% calibration errors have been added in quadrature to the fit errors.  The 'Class' column describes the source morphology: Jet = distinct jet, Jet? = possible jet.  A few known non-blazar sources are also indicated.}
\end{deluxetable}

\end{document}